\documentclass[notitlepage,twocolumn,prl,tightenlines,nofootinbib,superscriptaddress,showpacs]{revtex4}

\usepackage{amsmath}
\usepackage{amssymb,amsfonts}
\usepackage{bm}
\usepackage[mathcal]{euscript}
\usepackage{graphicx}
\usepackage{epsfig}
\usepackage{color}

\definecolor{blue}{rgb}{0.1,0.3,1}
\definecolor{green}{rgb}{0.1,0.6,0.1}
\definecolor{red}{rgb}{1,0,0}
\definecolor{pink}{rgb}{0.9,0.3,0.7}
\definecolor{azur}{rgb}{0,0.5,0.5}
\definecolor{orange}{rgb}{1,0.5,0.2}
\definecolor{brown}{rgb}{0.5,0,0}

\newcommand{\co}{(color online)~}

\begin{document}

\title{Evolution of dynamical facilitation approaching the granular glass transition}

\author{R. Candelier}
\affiliation{SPEC, CEA-Saclay, URA 2464 CNRS, 91 191 Gif-sur-Yvette, France}
\author{O. Dauchot}
\affiliation{SPEC, CEA-Saclay, URA 2464 CNRS, 91 191 Gif-sur-Yvette, France}
\author{G. Biroli}
\affiliation{Institut de Physique Th\'eorique, CEA, IPhT, F-91191 Gif-sur-Yvette, France and CNRS, URA 2306}

\begin{abstract}
We investigate the relaxation dynamics of a dense monolayer of bidisperse beads by analyzing the experimental data previously obtained in a fluidized bed. We show that the dynamics is formed by elementary relaxation events called cage jumps. 
These aggregate on a very short time into clusters. Increasing the packing fraction makes the spatio-temporal organization of the clusters evolve from a rather scattered and random distribution towards a collection of sparse and large events, called 
avalanches. The avalanche process is a manifestation of dynamical facilitation. The study of its 
evolution with density reveals that dynamical facilitation becomes less conserved and play a lesser role for the structural 
relaxation approaching the granular glass transition.
\end{abstract}

\maketitle


The dynamics of supercooled liquids~\cite{ediger1996sla,debenedetti2001sla}, colloids~\cite{weeks2000tdd} and agitated granular media~\cite{dauchot2007gbr} dramatically slows down as these systems approach the glass transition. Surprisingly, particles configurations close to the transition still look like the ones of a high temperature liquid. Instead, dynamical trajectories do show significant modifications. The motion becomes intermittent at the microscopic scale: typically a particle rattles for a long time inside a ``cage'' formed by its neighbours, before jumping into another ``cage''. Henceforth we shall call this event cage jump. Consecutive cage jumps lead to structural relaxation and long time diffusion. This phenomenon has been visually observed in colloids~\cite{weeks2002pcr}, granular media~\cite{pouliquen2003fpm,marty2005sac,candelier2009bbd} and numerical simulations of supercooled liquids~\cite{kudchadkar1995mds,candelier2010adc}. Another very important feature of glassy dynamics is the emergence of dynamical heterogeneity: there is by now experimental~\cite{ediger2000shd,berthier2005dee,dauchot2005dhc,lechenault2008csa,keys2007mgd,weeks2000tdd,cipelletti2005sdi} and numerical evidence~\cite{hurley1995kst,kob1997dhs} that dynamics becomes spatially correlated approaching the glass transition; there appear spatially localized regions relaxing much faster than the average. 
Providing a microscopic explanation for these phenomena has become a central issue in the field. Despite a number of theoretical proposals~\cite{tarjus2005fas,shintani2006fwc,lubchenko2006tsg, biroli2008tsg,garrahan2002prl,brito2007hdm}, 
there is still no consensus. One particularly debated question is the role of dynamical facilitation ($DF$) in glassy dynamics. $DF$ means that a local relaxation has a very high probability of happening nearby another relaxation after a certain time, which is short compared to the macroscopic relaxation time but large compared to the microscopic one. Effective models based on kinetic constrains \cite{fredrickson1984,garrahan2002prl} posit that $DF$ is the underlying cause of particle mobility by assuming that a region of jammed atoms can become unjammed and exhibit mobility only when it is adjacent to a region already unjammed.
Within the models this is due to the existence of mobility inducing defects, which cannot disappear (or appear) 
except if there is another defect nearby. This constraint implies that local relaxations cannot start or end without correspondingly being preceded or followed in space and time by other local relaxations. We will refer to this
property as { \it conservation} of $DF$. In other approaches \cite{wolynes2008fcg}, instead, $DF$ is an important piece of the theoretical description but not the driving mechanism of glassy dynamics.

Clearly, understanding how consecutive cage jumps conspire together and lead to macroscopic relaxation would be very instrumental in clarifying the role of dynamical facilitation and in explaining the emergence of dynamical heterogeneity. A first attempt in this direction has been made in the study of granular media~\cite{candelier2009bbd}, where we unveiled that dynamical heterogeneities arise from the aggregation of quasi-instantaneous clusters of cage jumps into long lasting avalanches. The dynamical process leading to avalanches provides a clear evidence of the important role played by $DF$: a local relaxation due to a cluster of cage jumps is typically followed nearby in space and in time by another cluster relaxation, and so on and so forth until the entire avalanche process is formed. In order to understand precisely the role played by $DF$, and to discriminate amongst the various theoretical scenarii, it is now crucial to characterize the evolution of the avalanche process when approaching the glass transition.


To this aim we focus on the $2D$ fluidized bed of beads studied in~\cite{keys2007mgd}, whose experimental data were generously provided by the authors.  The system is made of a 1:1 bidisperse mixture of $N$ steel beads of diameters $d_S = 0.318\textrm{cm}$ and $d_L = 0.397\textrm{cm}$ ($d_L/d_S=1.25$), with respective masses of $0.130\textrm{g}$ and $0.266\textrm{g}$, confined to a circular cell of diameter $17.7\textrm{cm} = 55.7d_S$. Bead motion is excited by an upward flow of air at a fixed superficial flow speed of $545\pm10\textrm{cm.s}^{-1}$ (resp. $500\pm10\textrm{cm.s}^{-1}$) for the $3$ loosest (resp. densest) packing fractions. The original acquisition frame rate is $120$~Hz; we retain one frame out of ten and follow the trajectories over $10,800$ frames. To avoid boundary effects, we consider a circular region of interest of diameter $D=45d_S$. All lengths are expressed in number of small grain diameters, and times in number of frames ($\frac{1}{12}\textrm{s}$). We study packing fraction ranging from $\phi=0.758$ to $\phi=0.802$ ($N=1,790$ to $1,975$). The data for the three most dense packings were not discussed in~\cite{keys2007mgd}.

\begin{figure}[t] 
\center
\includegraphics[width=0.49\columnwidth]{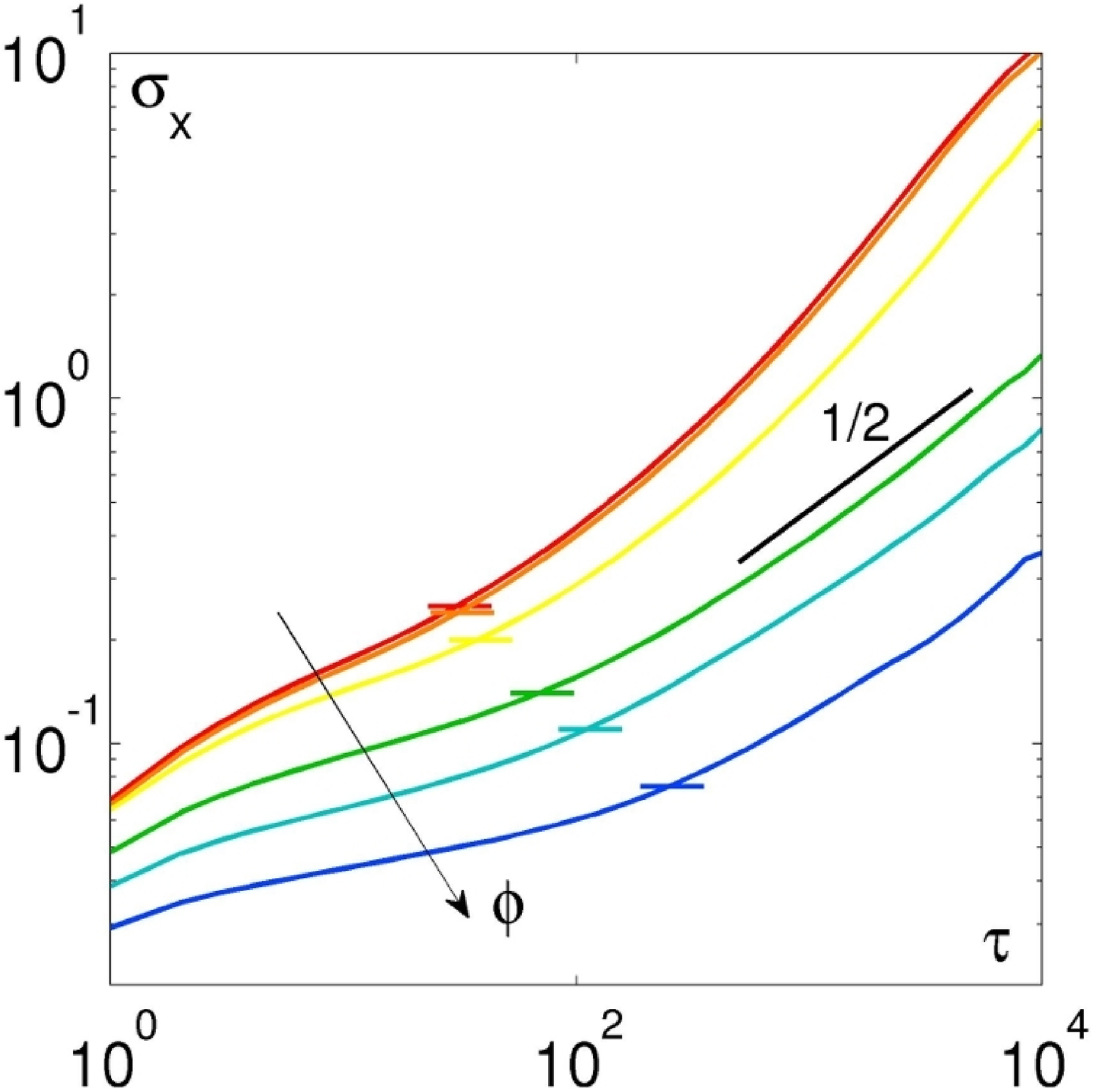}
\includegraphics[width=0.49\columnwidth]{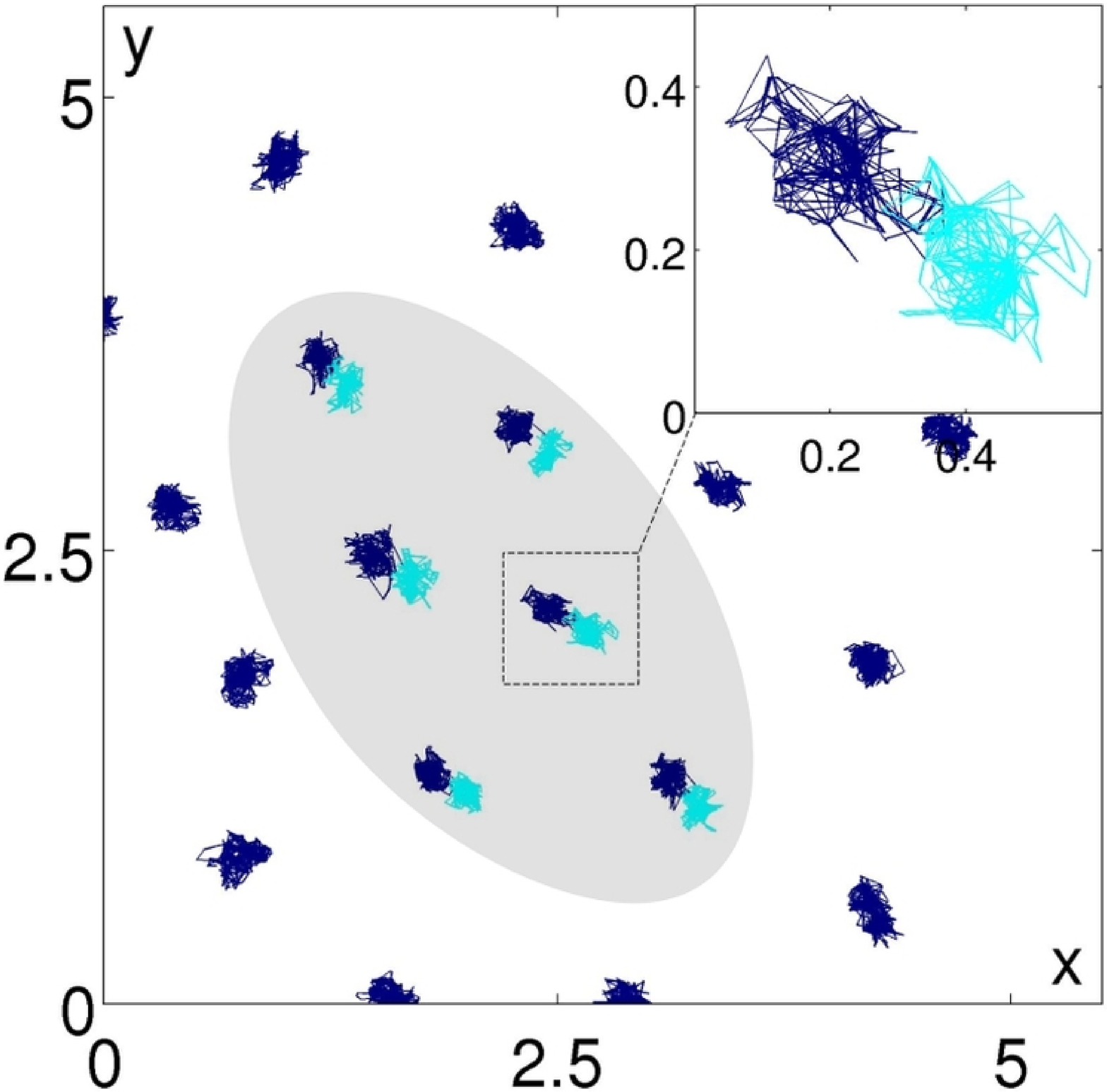}
\includegraphics[width=0.49\columnwidth]{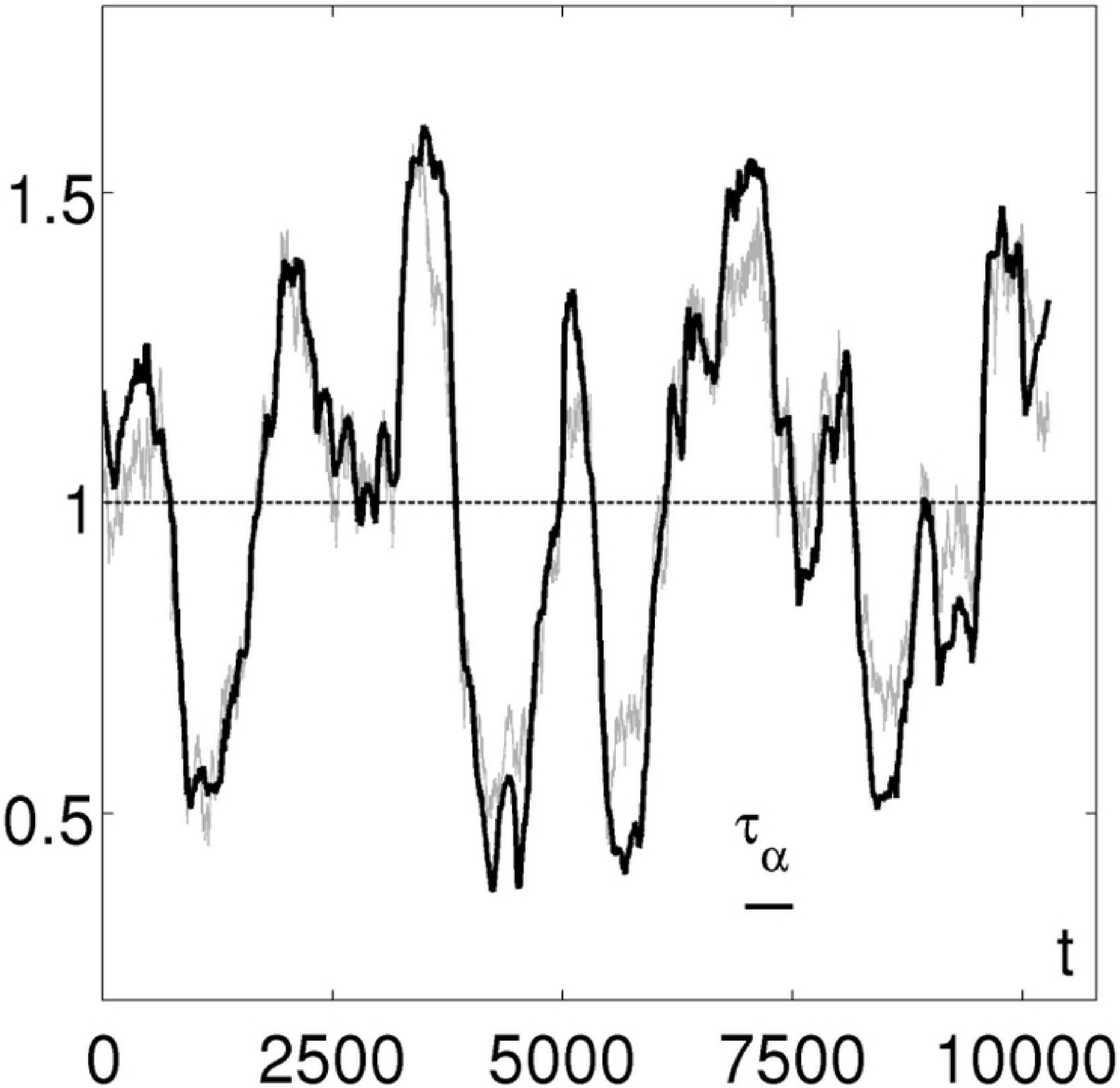}
\includegraphics[width=0.49\columnwidth]{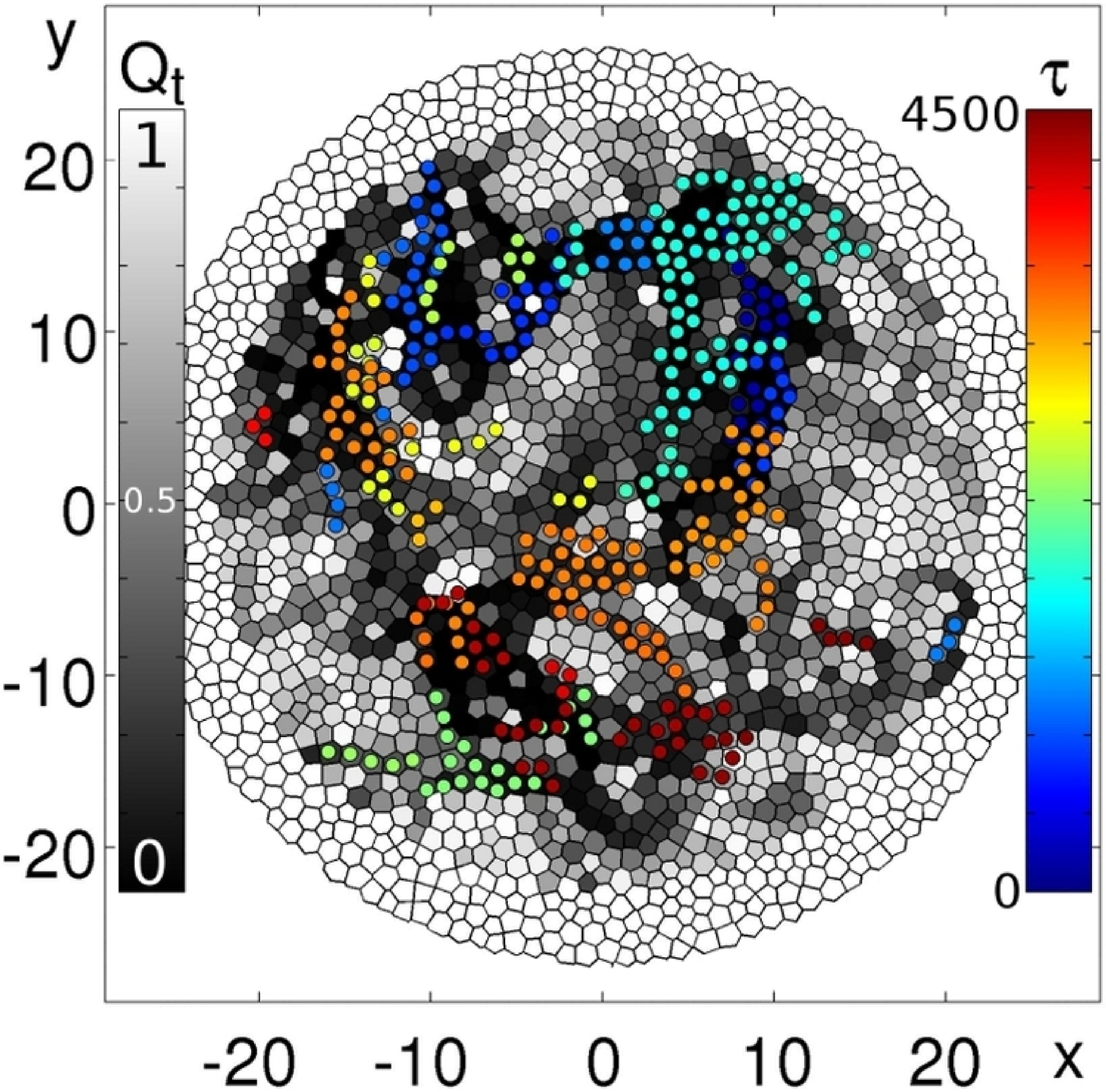}
\caption{\co Dynamics and cooperative jumps.
Top-left: Root mean square displacement along the $x$-direction $\sigma_x(\tau)$ for the $6$ packing fractions $\phi=0.758$ (red) to $0.802$ (blue). The horizontal dashes are located at the thresholds of the cage jump detection algorithm.
Top-right: Trajectories of a few particles at $\phi = 0.802$ for $1,000$ time frames. The color changes from blue (black) to cyan (light grey) when a cage jump is detected. All cage jumps in the grey area appear within $15$ time steps, defining a cooperative cluster.
Bottom-left: Comparison between the relative spatially averaged relaxation $Q_t(\tau_\alpha)/\langle Q_t \rangle_t$ (grey) and the relative percentage $P_t(\tau_\alpha)/\langle P_t \rangle_t$ of particles that haven't jumped between $t$ and $t+\tau_\alpha$ (black) at $\phi=0.773$, $\tau_\alpha=512$.
Bottom-right: Clusters of cage jumps appeared between $t$ and $t+\tau_\alpha$ superimposed on the map of $Q_{p,t}(\tau_\alpha)$ (grey-scale, left colorbar). The times $\tau$ at which clusters occur are color coded (right colorbar), $\phi=0.802$, $\tau_\alpha=4536$.
}
\label{fig:RMS_traj}
\vspace{-0.5cm}
\end{figure}

\begin{figure}[t] 
\center
\includegraphics[width=0.48\columnwidth]{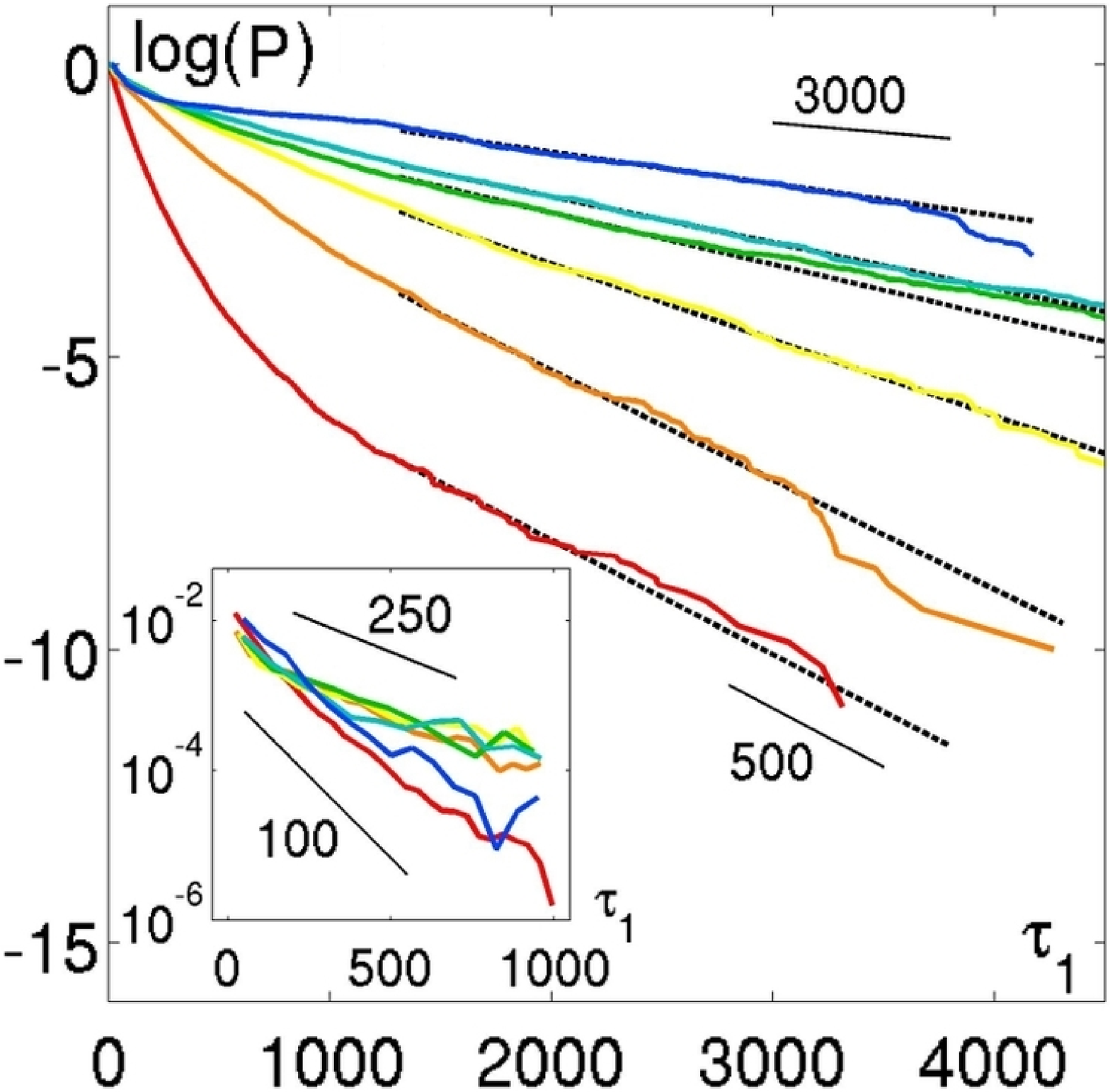}
\includegraphics[width=0.50\columnwidth]{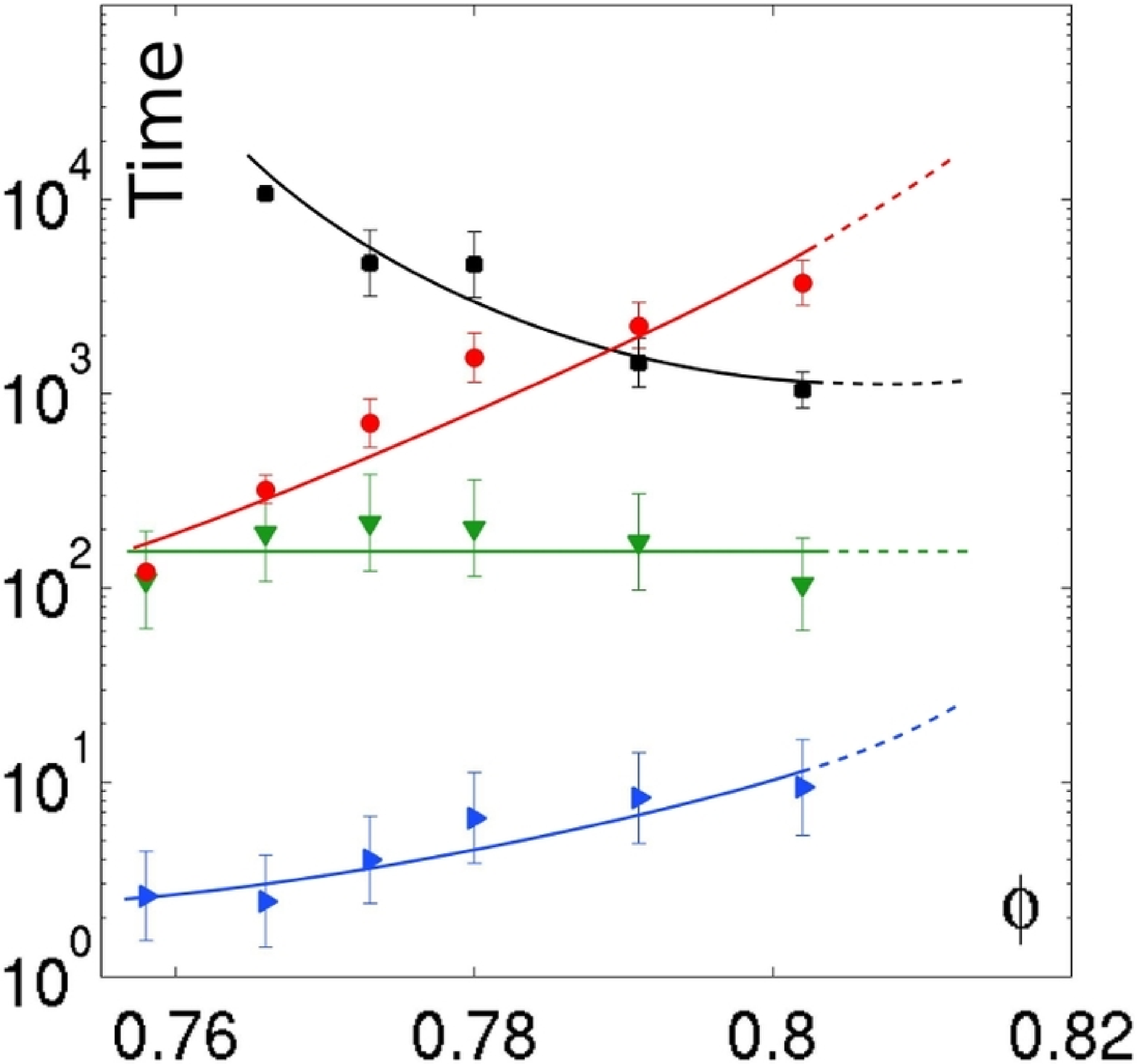}
\caption{\co Times scales as a function of packing fraction.
Left: Distributions of $\tau$, the lag time between adjacent clusters; \textit{Main plot} $P(\tau>\tau_1)$ -- black dotted lines are exponential fits at large $\tau_1$. \textit{Inset} Pdf$(\tau)$ for the population of short lag times (see text for more details) -- black lines are indicative exponential decays.
Right:  $\tau_{1/2}$~(\textcolor{red}{\textbullet}) , $\tau_{cl}$(\textcolor{blue}{$\blacktriangleright$}), $\tau_S$~(\textcolor{green}{$\blacktriangledown$}), and $\tau_f$~({\tiny$\blacksquare$}).
}
\label{fig:PDFs}
\vspace{-0.5cm}
\end{figure}

\begin{figure*}[t!] 
\centering
\includegraphics[width=0.62\columnwidth]{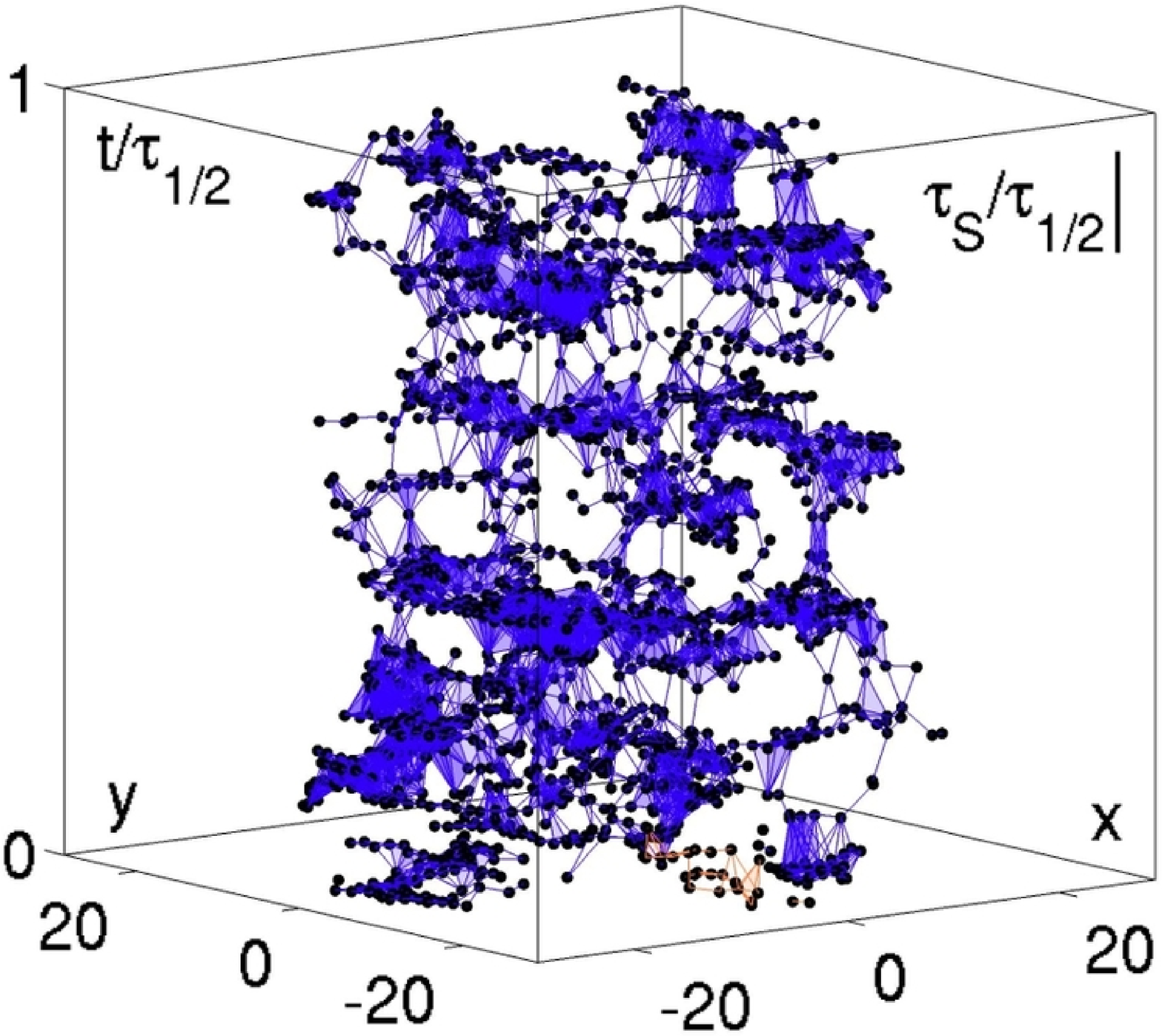}
\includegraphics[width=0.62\columnwidth]{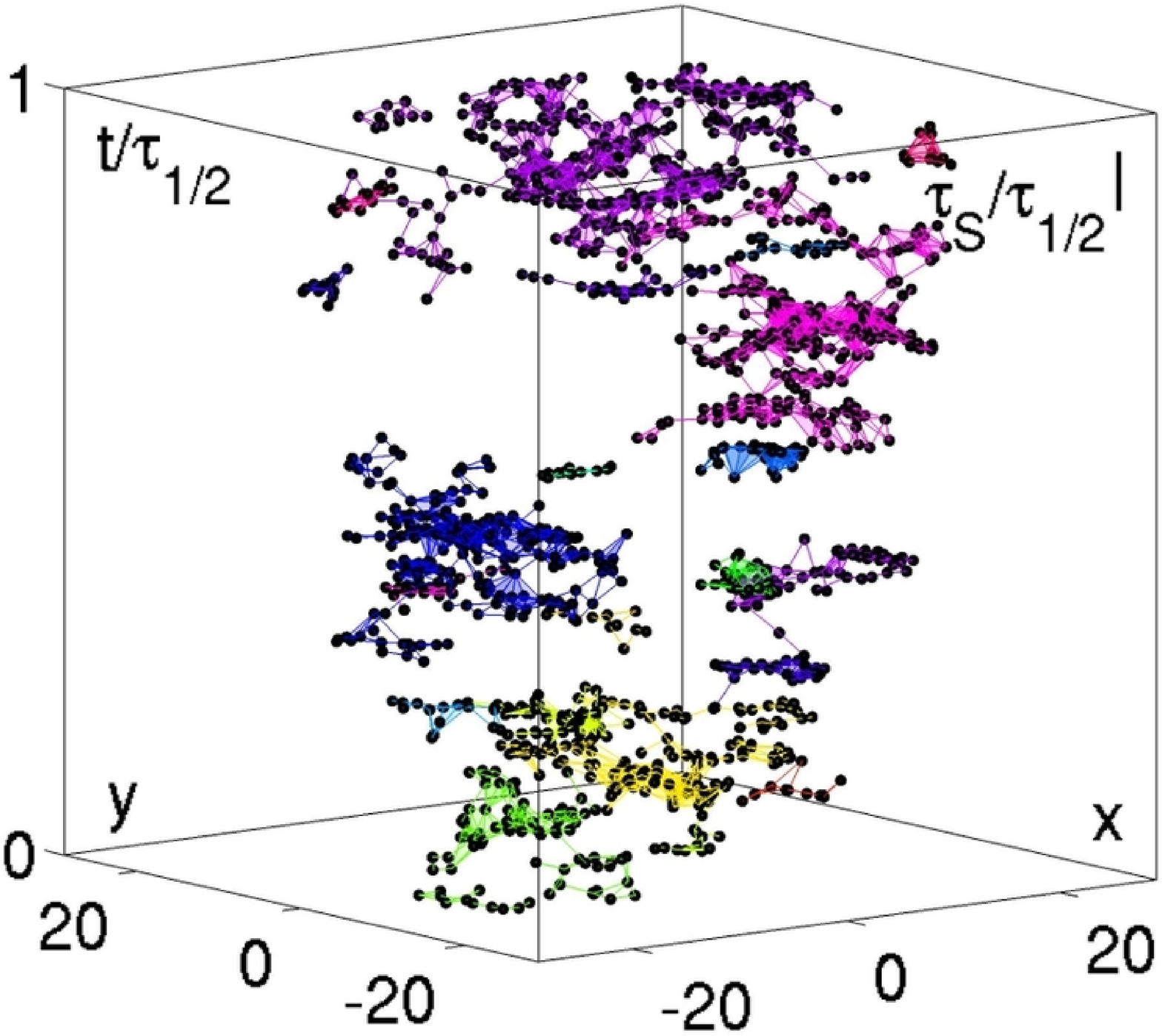}
\includegraphics[width=0.62\columnwidth]{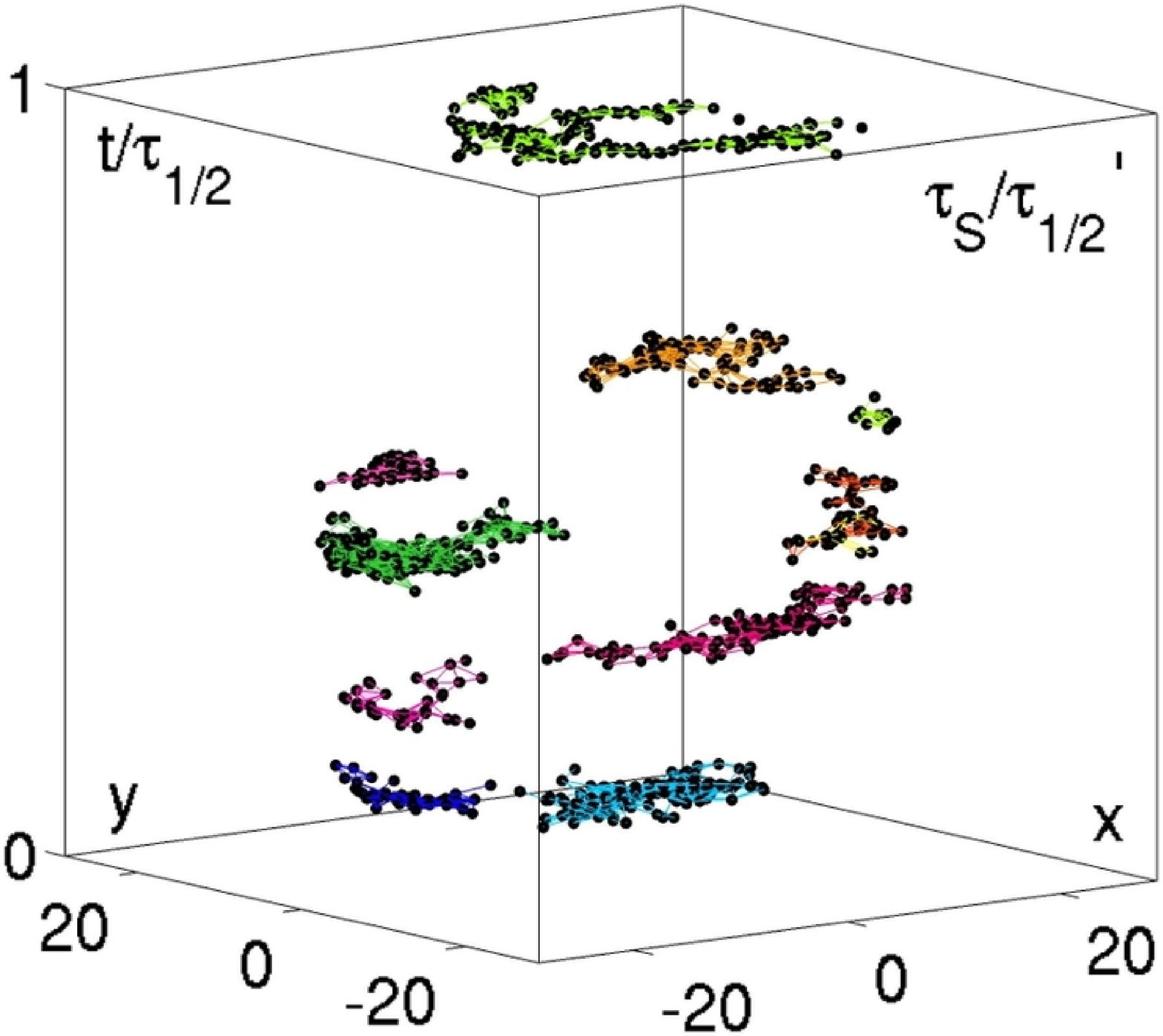}\newline
\hspace{14mm}
\includegraphics[width=0.55\columnwidth]{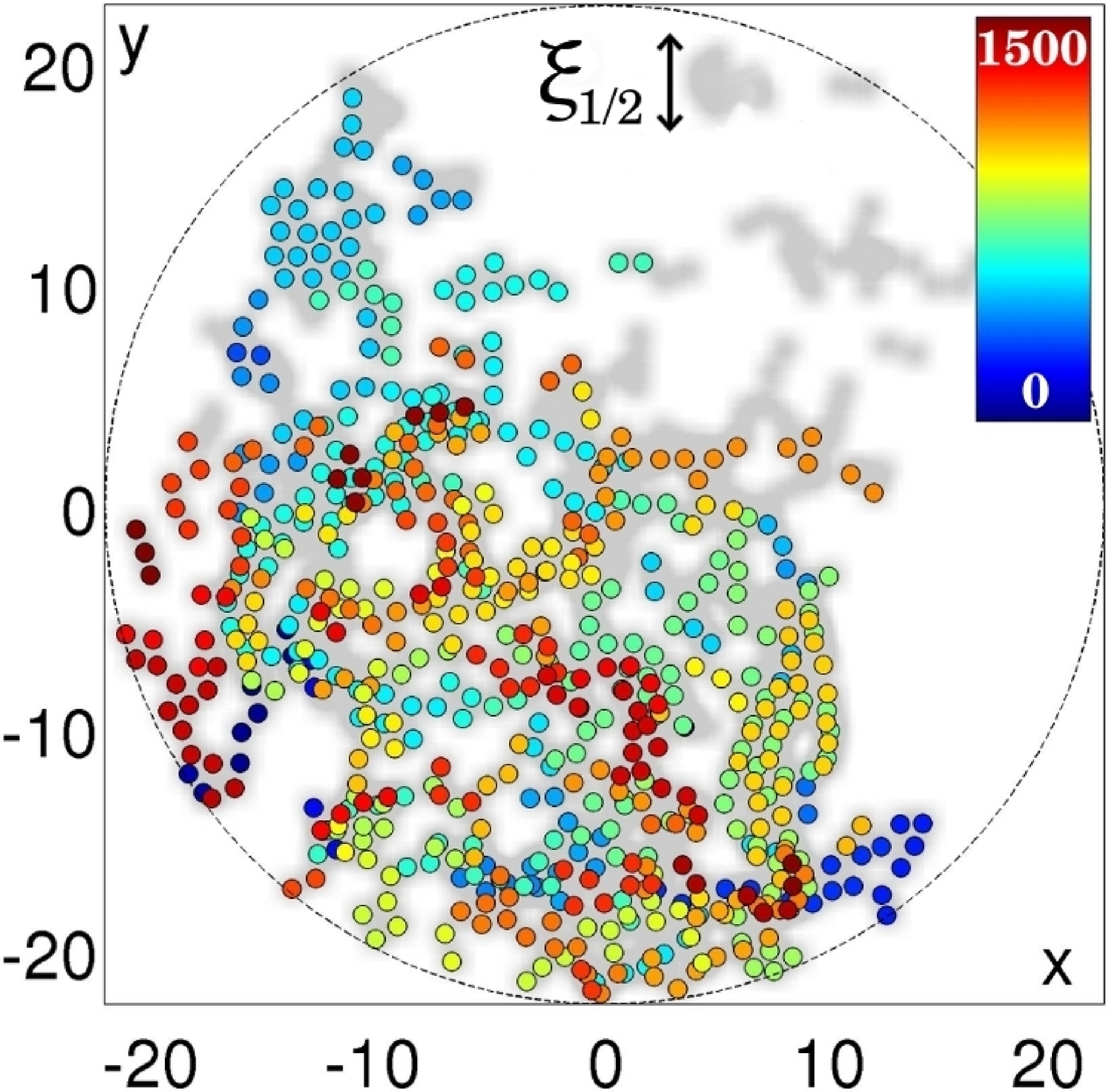}\hspace{10mm}
\includegraphics[width=0.55\columnwidth]{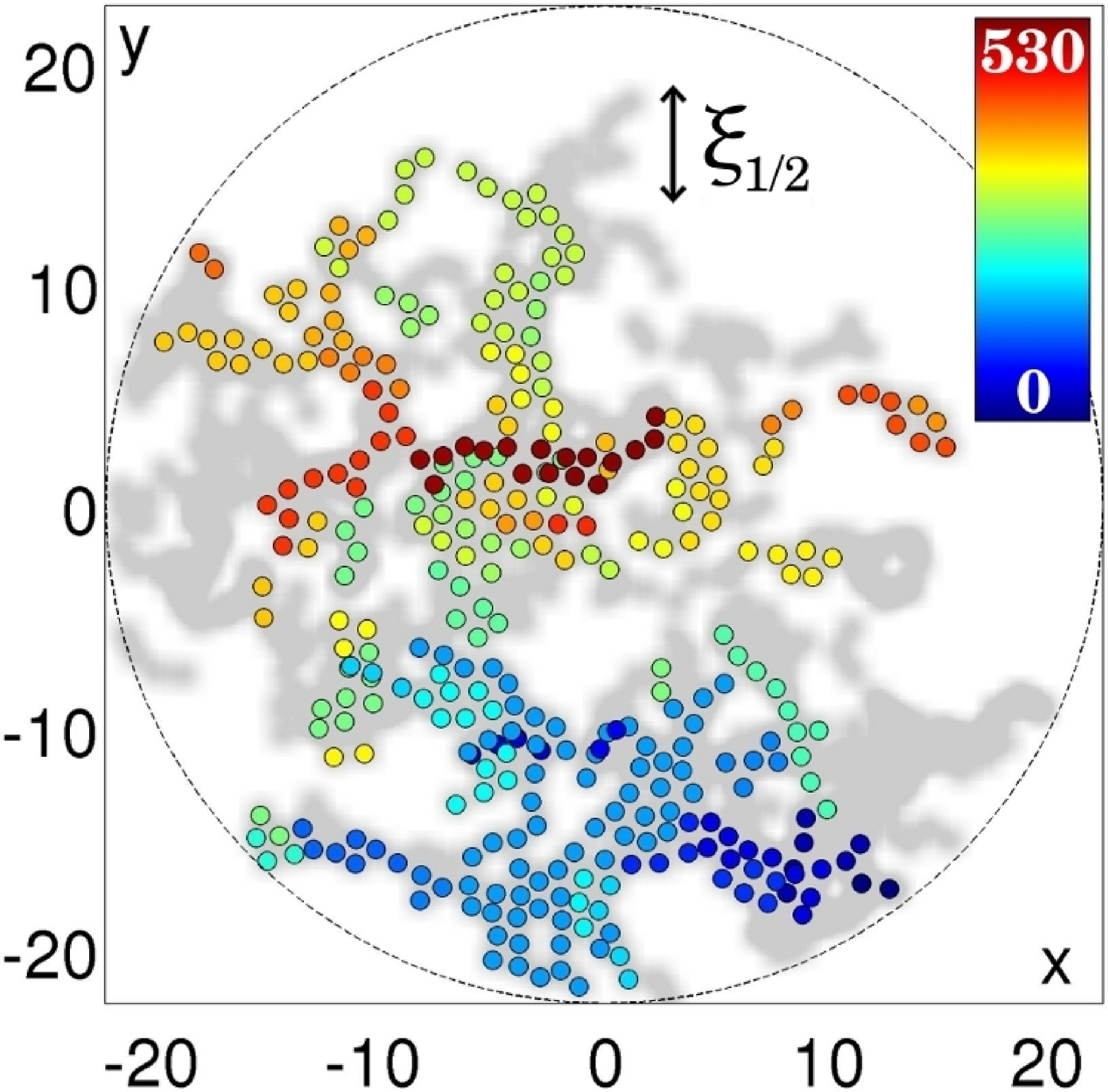}\hspace{10mm}
\includegraphics[width=0.55\columnwidth]{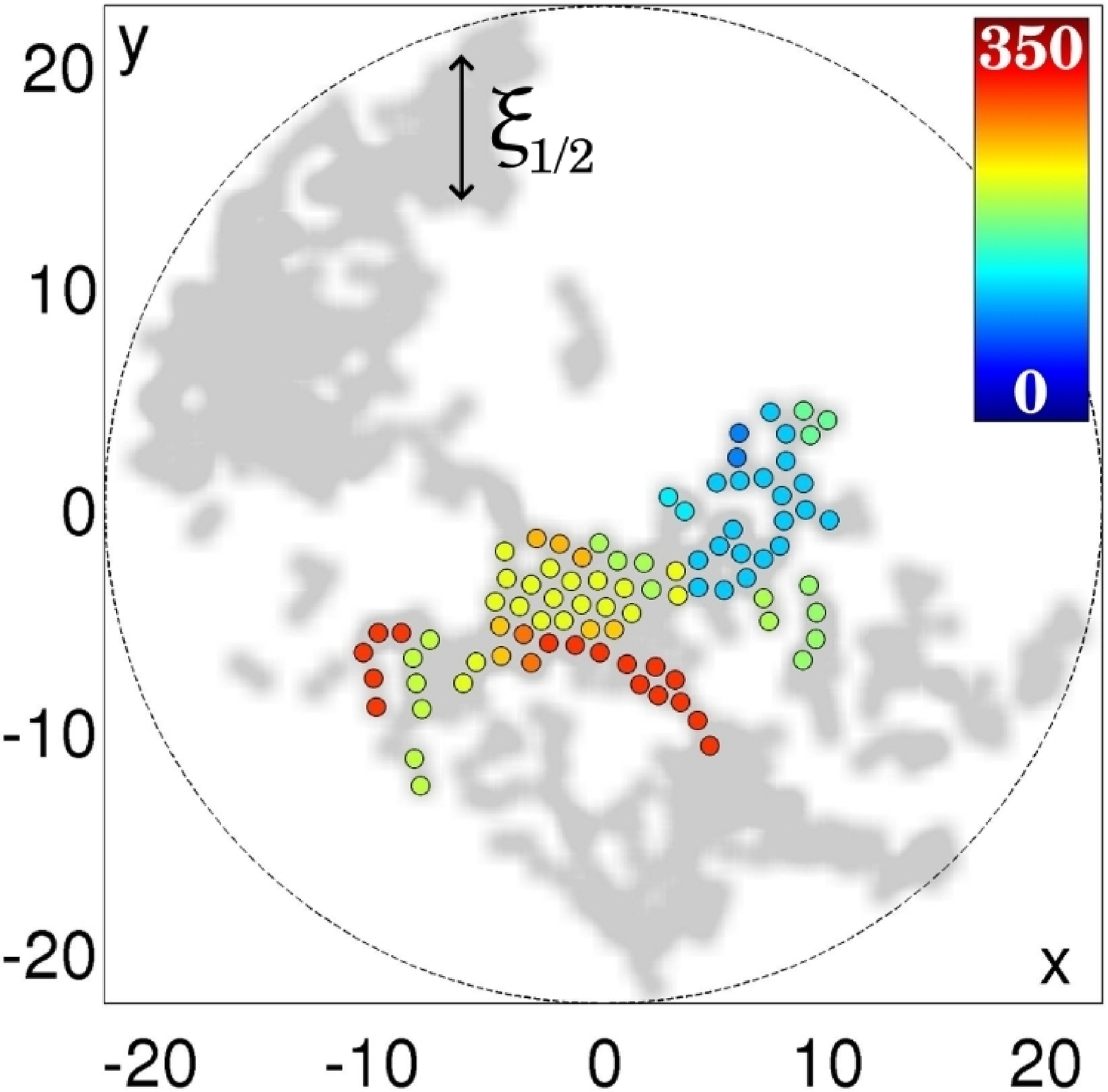}
\caption{\co
Top: Facilitation patterns in space and time during the typical relaxation time $\tau_{1/2}(\phi)$ for $3$ packing fractions~: from left to right $\phi=0.780$, $0.791$, $0.802$ and $\tau_{1/2} = 1540$, $2250$, $3730$. The two directions of space are in the horizontal plane and time is the vertical axis. The ratio $\tau_S/\tau_{1/2}$ is given in the upper-right corners. Jumps are represented with black dots, and all possible tetrahedrons which edges are the facilitating links between jumps are shown, forming volumes. Each separate connected structure has a different color.
Bottom: Jumps occurring in $\tau_{1/2}$ (in grey), same packing fractions. The jumps belonging to one arbitrarily chosen connected structure are colored according to the time at which they occur.
}
\label{fig:fp}
\vspace{-0.5cm}
\end{figure*}

In fig.~\ref{fig:RMS_traj} top-left we plot the root mean square displacement along the $x$-axis on a lag time $\tau$, $\sigma_x(\tau)$, which shows all the well-known characteristics observed when approaching the glass transition: a sub-diffusive plateau at intermediate time scales, which enlarges when increasing the packing fraction, and the final recovery of a diffusive regime on long times. For the three loosest packings the slope is greater than $1/2$, indicating the presence of slow convection rolls. This effect becomes stronger at even lower densities. Here we retain only the highest packing fraction, for which this does not interfere with the timescales of the analysis.

In order to analyze the microscopic relaxation processes, we apply the same procedure as developed in~\cite{candelier2009bbd} which allows one to obtain a coarse grained description of the dynamical evolution in terms of cage jumps: within a trajectory $S(t)_{t \in [0,T]}$, the time of the largest cage jump is given by the position $t_c$ of the maximum of $p(t) = \xi(t). [\langle d_1(t)^2 \rangle_{S_2}. \langle d_2(t)^2 \rangle_{S_1}]^{1/2}$, where $S_1$ and $S_2$ are the trajectory subsets $S\{t \in [0,t_c]\}$ and $S\{t \in ]t_c,T]\}$, $d_k(t)$ is the distance between the position at time $t$ and the center of mass of the subset $S_k$, the average $\langle . \rangle_{S_k}$ is computed over the subset $S_k$ and $\xi(t) = \sqrt{t_c/T(1-t_c/T)}$ is a natural statistical normalization. The procedure is repeated iteratively for every sub-trajectory until $\max(p)<\sigma_c^2$, where the thresholds $\sigma_c(\phi)$ are defined at the cross-over between subdiffusive and diffusive regimes in $\sigma_x(\tau)$ (see fig.\ref{fig:RMS_traj}-top left) and correspond to the sizes of a cage. This algorithm allows us to locate the cage jumps within a resolution of $15$ time steps (see fig.~\ref{fig:RMS_traj}-top right, which shows that cage jumps are well defined dynamical events). The evolution of the temporal correlation function is given by $Q_t(\tau) = \left< Q_{p,t}(\tau) \right>_p$ with
\vspace{-2mm}
\begin{equation*}
Q_{p,t}(\tau)=\exp \left( - \frac{||\Delta \vec{r}_p(t,t+\tau)||^2}{2a^2} \right),
\label{eq:def_Q}
\vspace{-2mm}
\end{equation*}
\noindent where $\Delta \vec{r}_p(t,t+\tau)$ is the displacement of particle $p$ between $t$ and $t+\tau$ and the length scale $a$ is set to $0.2$. Fig.~\ref{fig:RMS_traj}-bottom left shows that for $\tau=\tau_\alpha$, the relaxation time defined by $\left<Q_t(\tau)\right>_t=1/2$, $Q_t(\tau)$ is very well described by $P_t(\tau)$, the percentage of particles that have not jumped between $t$ an $t+\tau$.

Subsequent cage jumps aggregate into clumps that we call cooperative clusters and whose very short duration is denoted $\tau_{cl}$. This is clear from fig.~\ref{fig:RMS_traj}-bottom right, which shows that the large decorrelation patterns observed on time-scales $\tau_\alpha$ issue from the aggregation of several clusters of particles hopping at successive times.
In order to substantiate more quantitatively the existence of clusters we focus on the distribution of the lags $\tau$ separating the clusters that are adjacent in space and time. Figure~\ref{fig:PDFs}-top left displays $P(\tau>\tau_1)$, the probability of observing $\tau$ larger than $\tau_1$. As in~\cite{candelier2009bbd}, these cumulated distributions are well described by the addition of two processes : 
\vspace{-2mm}
\begin{equation*}
P(\tau>\tau_1) = \left(p_S e^{-\frac{\tau_1}{\tau_S}}+(1-p_S) e^{-\frac{\tau_1}{\tau_L}}\right);
\vspace{-2mm}
\end{equation*}
$p_S$ is the fraction of short lag times. The short time scale, $\tau_S$, physically corresponds to dynamic facilitation events: cluster relaxations followed closely in time and in space by other cluster relaxations. The long time scale $\tau_L$ corresponds to the average time spent in a cage. Technically, we extract first $\tau_L$ and $p_S$ by fitting the large $\tau_1$ regime, then we subtract the large $\tau_1$ contribution and obtain the exponential distribution for the short lag times displayed in the inset of fig.~\ref{fig:PDFs}-top left and from which one easily estimates $\tau_S$.

We now come to the central discussion of this work: the evolution of the above dynamical patterns, identical to the ones observed in~\cite{candelier2009bbd}, when the packing fraction is increased towards the glass transition.
One observes on fig.~\ref{fig:PDFs}-top left that the relaxation time $\tau_\alpha$ or its alternative estimation $\tau_{1/2}$, the time needed for observing half of the particles to jump once, increases strongly with the packing fraction, while the cooperative clusters typically last a short time $\tau_{cl}$ varying from $2$ to $10$, not a significant variation given our temporal resolution on the detection of the cage jumps. $\tau_S$ remains bounded between $100$ and $250$ without clear tendancy, while $\tau_L$ increases from $511$ to $3,041$, following the slowing down of the dynamics. Note that $\tau_S$ is larger than $\tau_{cl}$ thus confirming that clusters are well defined dynamical events.

Clearly, the picture of clusters dynamically facilitating each others only makes sense when $\tau_L$ becomes larger than $\tau_S$, that is above $\phi^*\sim0.77$, which would be analogous to the onset temperature in supercooled liquids. 
The way in which clusters aggregate and the resulting facilitation patterns are represented in fig.~\ref{fig:fp}-top for three packing fractions in 3D space/time, the time axis being rescaled with respect to the relaxation time $\tau_{1/2}$. We draw all cage jumps (black dots) and link the ones separated by a lag time less than $\tau_S$. This defines a network whose vertices are the cage jumps and whose edges are the orientated links towards facilitated jumps. 
For the loosest packing fraction, all jumps are connected by a facilitation link and form a highly interconnected monolith where facilitation appears to be conserved. When raising $\phi$, an increasing number of adjacent clusters become separated by more than a few $\tau_S$ within a time interval equal to the relaxation time (the ratios $\tau_{1/2}/\tau_S$ and $\tau_L/\tau_S$ increase up to $20$ and $30$).  Eventually, the facilitation network does not percolate in time anymore and separated avalanches form. 
The average duration of the avalanches, that we call facilitation time, $\tau_f$, decreases and becomes smaller than $\tau_{1/2}$ for the largest $\phi$. At that point facilitation is clearly not conserved anymore. In agreement with the above discussion, we find that $p_S$ decreases from $90\%$ to $40\%$, suggesting that facilitation occurs for a decreasing number of clusters. It would be interesting to check whether at even higher
density $\tau_f$ becomes of the order of $\tau_S$: each avalanche would reduce to a single cluster and facilitation would disappear completely.

\begin{figure}[t]
\center
\includegraphics[width=0.49\columnwidth]{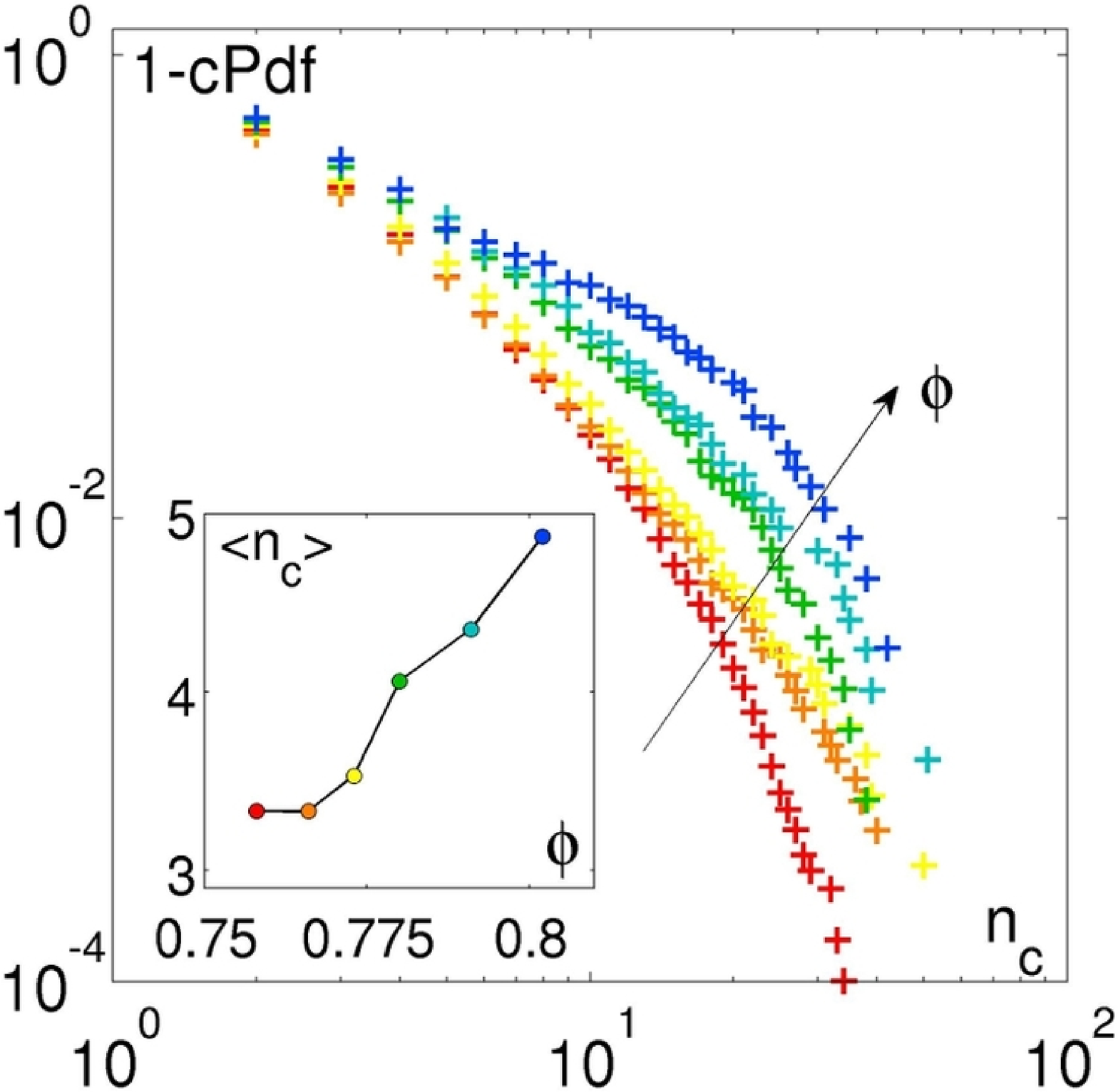}
\includegraphics[width=0.47\columnwidth]{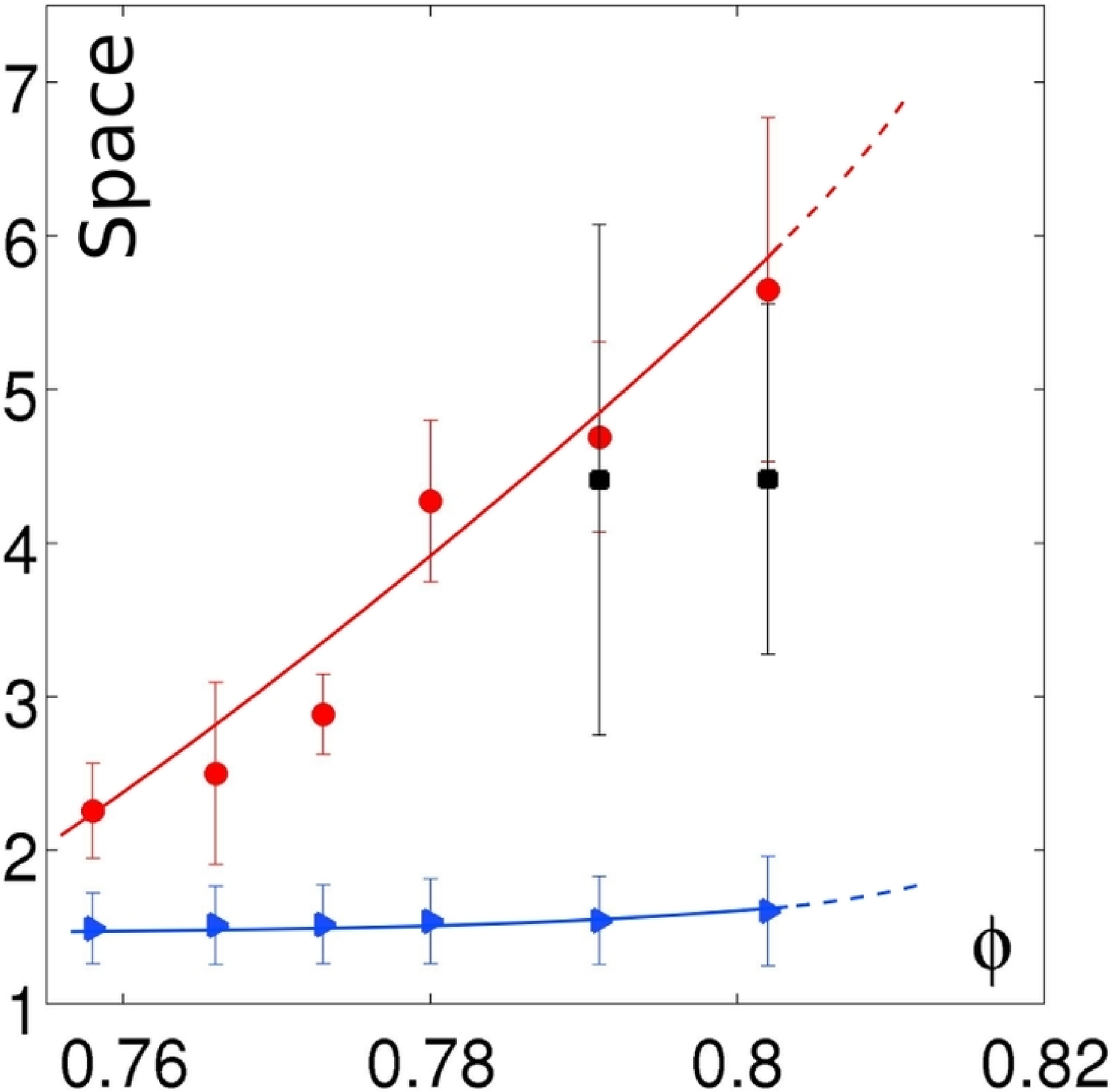}
\caption{\co
Length scales as functions of the packing fraction.
Left: Cumulated Pdf of the clusters' number of particles $n_c$ for the $6$ packing fractions $\phi=0.758$ (red) to $0.802$ (blue); \textit{Inset} Mean value of $n_c$ over all clusters, as a function of $\phi$.
Right: $\xi_{cl}$(\textcolor{blue}{$\blacktriangleright$}), $\xi_{1/2}$~(\textcolor{red}{\textbullet}), and $\xi_{ava}$~({\tiny$\blacksquare$}).
Plain lines are guides for the eyes, dashed lines are extrapolations.
}
\label{fig:timelength_scales}
\vspace{-0.5cm}
\end{figure}

Finally, we address the evolution of the dynamical correlation length-scale $\xi_{1/2}$ estimated as the correlation length of the facilitation pattern (assuming by ergodicity that spatial and time averages coincide). $\xi_{1/2}$ 
is approximatively equal to the the average width of the backbone forming the pattern. 
Figure~\ref{fig:fp}-bottom displays the spatial projection of cage jump during $\tau_{1/2}$. In grey, one sees the half of the particles which have jumped and we have colored those belonging to one arbitrarily chosen connected structure. For the lowest $\phi$ (left panel) $\tau_f>\tau_{1/2}$ and almost all cage jumps belong to the same large, eventually infinite connected structure. $\xi_{1/2}$ is roughly the cluster size, thus showing that the pattern is formed by dynamically independent clusters. In this regime, the pattern is so much intertwined that a clusters is facilitated by several others. Thus, dynamical correlations do not propagate farther than the size of one single cluster.  At higher $\phi$s, the distributions of the cluster sizes $n_c$ (see fig.~\ref{fig:timelength_scales}-top left) have larger tails and their experimental average $\langle n_c \rangle$ grows from $3.4$ to $5$. For the packing fraction corresponding to the middle panel of Fig.~\ref{fig:fp}-bottom, $\tau_f \simeq \tau_{1/2}$, the clusters are slightly larger and more concentrated and $\xi_{1/2}$ is increased, as shown in fig.~\ref{fig:timelength_scales}-top right. Finally, at the highest density (right panel of  Fig.~\ref{fig:fp}-bottom), when avalanches are well formed and separated, we find that $\xi_{1/2}$ is again increased and has become of the order of the avalanche size $\xi_{ava}$. At this packing fraction, $\tau_f < \tau_{1/2}$ and the clusters are even more grouped. 

To summarize, analyzing data coming from a fluidized monolayer experiment, we have confirmed the role of the spatio-temporal organization of cage jumps in the relaxation dynamics previously pointed out in our cyclic shear experiment. In both cases such cage jumps occur in cooperative clusters which give rise to facilitation. Above a characteristic packing fraction, akin to the onset temperature in liquids, facilitation starts to play a role in the dynamics. Increasing the packing fraction facilitation patterns evolve from a single connected structure percolating in time to isolated denser avalanches of finite duration. Dynamical correlations are, at first, of the size of clusters and then, in the latter regime, are of the size of the avalanches. 
Finally, approaching the granular glass transition, the cluster size increases whereas the number of 
facilitated clusters inside an avalanche decreases. Thus the cooperative relaxation of the first cluster of an avalanche plays a larger role and facilitation a lesser one. Investigating whether our findings also hold for supercooled liquids would certainly be of great interest. 


We thank A.R. Abate and D.J. Durian for sharing their data and J.-P. Bouchaud, D.R. Reichman and L. Berthier
for feedback on this manuscript.  This work was supported by ANR DYNHET 07-BLAN-0157-01.


\vspace{-7mm}
\bibliography{biblio_glass}

\end{document}